\documentstyle[aps,twocolumn,epsfig]{revtex}

\newcommand{\beq}{\begin{equation}}
\newcommand{\eeq}{\end{equation}}
\newcommand{\bea}{\begin{eqnarray}}
\newcommand{\eea}{\end{eqnarray}}
\newcommand{\ba}{\begin{array}}
\newcommand{\ea}{\end{array}}
\newcommand{\bc}{\begin{center}}
\newcommand{\ec}{\end{center}}
\newcommand{\lsimeq}{\stackrel{<}{\scriptstyle\sim}}
\newcommand{\gsimeq}{\stackrel{>}{\scriptstyle\sim}}

\newcommand{\etal}{{\it et al.}}
\newcommand{\bml}{\begin{mathletters}}
\newcommand{\eml}{\end{mathletters}}

\newcommand{\commentout}[1]{{}}
\newcommand{\half}{\hbox{$1\over2$}}

\newcommand{\HD}{{\cal H}}

\newcommand{\K}{{\cal K}}

\newcommand{\phidag}{\phi^\dagger}
\newcommand{\psidag}{\psi^\dagger}
\newcommand{\adag}{a^\dagger}
\newcommand{\bdag}{b^\dagger}
\newcommand{\cdag}{c^\dagger}

\newcommand{\r}{{\bf r}}

\newcommand{\k}{{\bf k}}

\newcommand{\eq}[1]{(\ref{#1})}
\newcommand{\vol}[1]{{\bf #1}}

\newcommand{\rbpot}{$^{87}\rm{Rb}\,$-$^{40}$K }

\begin{document}
\flushbottom
\draft
\wideabs
{
\title{Shortcut to a Fermi-Degenerate Gas of Molecules via Cooperative
Association}

\author{Olavi Dannenberg,$^1$ Matt Mackie,$^1$ and Kalle-Antti
Suominen$^{1,2}$}
\address{$^1$Helsinki Institute of Physics, PL 64, FIN-00014
Helsingin yliopisto, Finland \\
$^2$Department of Physics, University of Turku, FIN-20014 Turun yliopisto,
Finland}
\date{\today}

\maketitle

\begin{abstract}
We theoretically examine the creation of a Fermi-degenerate gas of
molecules by considering a photoassociation or Feshbach resonance applied
to a degenerate Bose-Fermi mixture of atoms. This problem raises an
interest because, unlike bosons, fermions in general do not behave
cooperatively, so that the collective conversion of, say, two million
atoms into one million molecules is not to be expected. Nevertheless, we
find that the coupled Fermi system displays collective Rabi-like
oscillations and adiabatic passage between atoms and molecules, thereby
mimicking Bose-Einstein statistics. Cooperative association of a
degenerate mixture of Bose and Fermi gases could therefore serve as a
shortcut to a degenerate gas of Fermi molecules.

\end{abstract}

}

\narrowtext

Laser cooling of molecules is difficult, if
not impossible, to apply in practice. This bottleneck occurs because the
removal of energy from the translational degrees of
freedom is most often
accompanied by excitation of ro-vibrational
modes which, upon de-excitation, couple back into the translational
motion in a manner that heats the system. A laser-cooling-based
approach\cite{DEM99} to a Fermi-degenerate gas of molecules is therefore
unlikely to succeed. Moreover, buffer-gas cooling~\cite{WEI98} and
Stark-deceleration~\cite{BET00} techniques, while no doubt of practical
interest in their own right, currently view quantum degeneracy from the
horizon.

One possible route to ultracold molecules is photoassociation\cite{WEI99},
and the formation of nondegenerate Fermi molecules was indeed
recently observed\cite{SCH02}. A shortcut to quantum degenerate
molecules is in the early stages of development for Bose-Einstein
condensates (BECs), so far delivering strongly enhanced molecule
formation\cite{WYN00}, precise measurements of the light-induced shift of
the binding energy\cite{GER01}, and searches for a fundamental limit to
the atom-molecule conversion rate\cite{MCK02,PIC02}. The magnetic field
counterpart to photoassociation is the Feshbach resonance\cite{TIM99a}.
Often seen as a tool for adjusting the scattering lengths, thereby
enabling condensation in otherwise-incondensible
systems\cite{COR00,WEB02}, the latest studies of a BEC tuned near a
Feshbach resonance have culminated in evidence for atom-molecule
coherence\cite{DON02}.

Motivated by these exciting experiments, we report that a degenerate gas
of Fermi molecules could be formed by applying either a
photoassociation~\cite{SCH02,WYN00,GER01,MCK02,PIC02} or Feshbach
resonance~\cite{COR00,WEB02,DON02} to a degenerate mixture of Bose and
Fermi atoms~\cite{TRU01,SCH01,MOD02}. The collective Fermi atom-molecule
system is predicted to undergo Rabi-like oscillations, as well as
adiabatic passage, from atoms to molecules, thereby mimicking cooperative
behavior that was previously attributed to Bose
statistics~\cite{MAC00,JAV02}.

Photoassociation occurs when two atoms absorb a laser photon, thereby
jumping from the free two-atom continuum to a bound molecular
state. If the initial atoms form a Bose-Einstein condensate, then a
photoassociation laser could conceivably be used to convert
a BEC of atoms into a BEC of molecules~\cite{JAV99,HEI00,VAR01}.
Collective free-bound photoassociation is theoretically identical to
magnetoassociation, whereby a molecular condensate can be created when
one atom from a condensed pair spin flips in the presence of a magnetic
field tuned near a Feshbach resonance~\cite{vABE99,TIM99,YUR99}.
Intuition developed in one instance is therefore generally applicable to
the other, and we will often refer simply to collective association.

Analogous with coherent optical transients in few level atomic
systems~\cite{ALL87}, photoassociation of a BEC has been predicted to
induce Rabi-like oscillations between atomic and molecular
condensates~\cite{JAV99,HEI00,VAR01}, whereby an entire gas of, say,
two million Bose-condensed atoms are collectively converted into a million
molecules that are, in turn, collectively converted back into (roughly)
two million atoms, {\it ad infinitum}. Another interesting possibility
arises because the ground state of the system is all atoms for large
positive detunings (far below threshold) and all molecules for large
negative detunings (far above threshold), so that a slow sweep of the
laser detuning from one extreme to the other will collectively convert a
BEC of atoms into a BEC of molecules~\cite{JAV99}. Incidentally, it was a
particular combination of these two concepts, applied instead to
magnetoassociation, that led to the observation\cite{DON02} of collective
Ramsey fringes between an atomic condensate and a small fraction of
molecular condensate dressed by dissociated atom
pairs~\cite{KOK02,MAC02,KOE03}.

The statistics of neutral atoms is determined
by the number of neutrons in the nucleus, which must be odd for fermionic
atoms and even for bosonic atoms. The sum of the total number of
neutrons in the nucleus of the constituent atoms similarly determines the
statistics of neutral molecules. Accordingly, molecules formed by
free-bound association of two fermions will necessarily result in a
boson, whereas fermionic molecules are born from the union of a boson and
a fermion.  Given a degenerate mixture of Bose and Fermi
gases~\cite{TRU01,SCH01,MOD02}, is it possible that collective free-bound
association could serve as a source of degenerate Fermi molecules?

To address this question, we model
a degenerate Bose-Fermi mixture of atoms~\cite{TRU01,SCH01,MOD02}
coupled by either a Feshbach or photoassociation resonance to a
Fermi-degenerate gas of molecules. The initial bosonic [fermionic] atoms
are denoted by the field
$\phi(\r,t)$ [$\psi_-(\r,t)$], and the fermionic molecules by the field
$\psi_+(\r,t)$. Neglecting particle-particle interactions, the Hamiltonian
density for such an untrapped system can be written
$\HD=\HD_0+\HD_I$, where
\bml
\bea
{\HD_0\over\hbar} &=&
\phidag\left[-{\hbar\nabla^2\over2m_B}-\mu_B\right]\phi
  +\psidag_-\left[-{\hbar\nabla^2\over2m_-}-\mu_-\right]\psi_-
\nonumber\\&&
  +\psidag_+\left[-{\hbar\nabla^2\over2m_+}+\delta_0-\mu_+\right]\psi_+,
\\
{\HD_I\over\hbar} &=& -\half {\cal K} \left(\psidag_+\psi_-\phi
  +\phidag\psidag_-\psi_+\right).
\eea
\label{FULL_HD}
\eml
Here $m_B$ ($m_-$) is the mass of a bosonic (fermionic) atom,
$m_+=m_B+m_-$ is the mass of a molecule,
$\mu_{B(\pm)}$ is the so-called chemical potential that implicitly
accounts for particle trapping,
$\delta_0$ is the binding energy (detuning) of the molecular state
relative to the dissociation threshold,
and ${\cal K}$ is the (real) atom-molecule coupling.

We now make some simplifications to allow for ease of modeling. First,
we scale the Fermi (Bose) atom fields as
$\psi_-\rightarrow e^{-i\mu_-t}\psi_-$
($\phi\rightarrow e^{-i\mu_Bt}\phi$), and the molecule field as
$\psi_+\rightarrow e^{-i(\mu_-+\mu_B)t}\psi_+$, which yields
$\mu_+\rightarrow \mu'_+=\mu_+-(\mu_-+\mu_B)$; in turn, $\mu'_+$ is
absorbed into the detuning. Second is that atom-molecule conversion occurs
on a timescale much faster than the motion of the atoms in the trap,
allowing us to neglect the kinetic energies, and justifying our omission
of an explicit trapping potential for the particles. Third
is that, whatever the Fermi energy associated with the fermionic
components, it lies within the Wigner threshold regime, so
that the coupling
$\K$ can be taken as the same for all modes. Finally, we switch to
momentum space, but retain only the $\k=0$ atomic condensate modes since,
due to Bose stimulation, these transitions are favored over $\k\neq 0$
modes. The simplified Hamiltonian reads
\beq
H=\sum_\k\left[\delta\cdag_\k c_\k
  -\half\kappa (\cdag_\k b_\k a +\adag\bdag_\k c_\k)\right],
\label{SIMP_HAM}
\eeq
where $a$ annihilates a bosonic atom with
wavevector \mbox{$\k=0$}, $b_\k$ ($c_\k$) annihilates an
atom (molecule) with wavevector $\k$, and $\kappa={\cal K}/\sqrt{V}$.

The system described by the Hamiltonian~\eq{SIMP_HAM} is a particular
nonlinear version of a two-level atom driven by an external
field. Of course, the two-level atom is well known to
undergo Rabi flopping~\cite{ALL87}, and similar behavior observed for the
present system would in principle answer question of
whether collective association will provide a shortcut to the formation
of a Fermi-degenerate gas of molecules. We therefore consider an
on-resonance system, for which
$\delta=0$ and the Heisenberg equations of motion are
\bml
\bea
&\displaystyle{i\dot{a} = \kappa \sum_\k\bdag_\k c_\k,}& \\
&\displaystyle{i\dot{b}_\k = \kappa \adag c_\k,}& \\
&\displaystyle{i\dot{c}_\k = \kappa b_\k a.}&
\eea
\label{HEIS_EQ}
\eml
If the system starts out with $N_B$ bosonic atoms, the operator $a$ will
have a characteristic size $\sqrt{N_B}$. On the other hand, while the
fermionic operators $b_\k$ and $c_\k$ are characteristically of order unity,
the sum over momentum states will ultimately introduce the initial number
of fermionic atoms. Hence, for equal numbers of bosons and fermions,
$N_B=N_F=N$, it is intuitively obvious that the equations of
motion~(\ref{HEIS_EQ}) yield a system that evolves with the characteristic
frequency $\Omega=\sqrt{N}\kappa$.

\begin{figure}
\centering
\epsfig{file=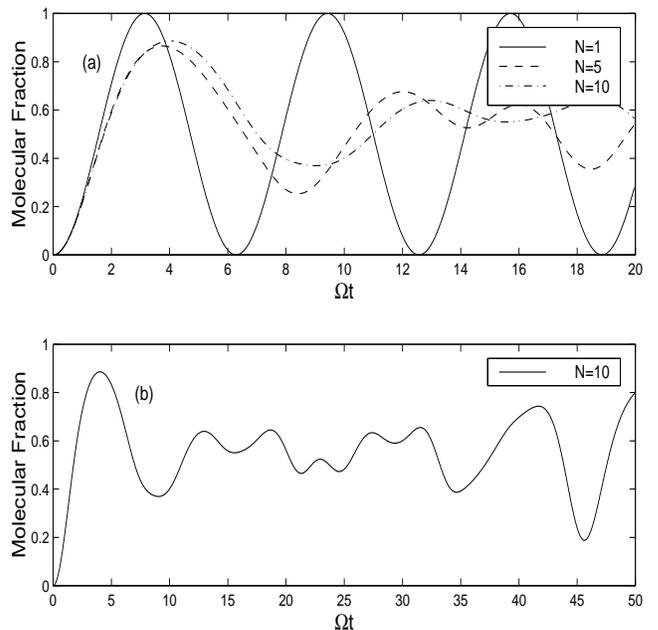,width=8.5cm,height=8.5cm}
\caption{Rabi-like oscillations in the fraction of Fermi-Bose atoms
converted to Fermi molecules.
The initial number of bosonic and fermionic atoms are equal, i.e.,
$N_B=N_F=N$. (a) The oscillations are complete for an initial
number of atoms
$N=1$, while for
$N=5$ many-body effects lead to frequency-shifted oscillations that are
incomplete and collapse. Better short-time agreement with the $N=1$ result
is obtained for $N=10$. (b) The oscillations eventually
revive.}
\label{RABI}
\end{figure}

Using Fock states, said intuition is confirmed by solving the
time-dependent Schr\"odinger equation numerically, the results of which
are shown in Fig.~\ref{RABI}. For the simplest case of
$N=1$, the system undergoes complete conversion from a doubly-degenerate
Fermi-Bose system of atoms to a Fermi-degenerate gas of molecules in a
time $\Omega^{-1}$. However, considering a larger number of initial
atoms, $N=5$, we see that quantum many-body fluctuations not only
frustrate complete conversion, but also adjust the oscillation frequency
and lead to collapse and revival. Increasing the initial particle number
to $N=10$ brings the amplitude of the first half-oscillation closer to
the ideal case. This behavior is exactly analogous to the
single-component bosonic case~\cite{JAV99,VAR01}. Although limited
computational resources preclude the explicit investigation of large
particle number, based on the bosonic analogy~\cite{JAV99,VAR01} we fully
expect the first half oscillation to be complete for large $N$, i.e.,
$N\gg 1$ initial Fermi atoms should be converted to $N$ Fermi molecules
over a timescale $\tau\sim\Omega^{-1}$.

Nevertheless, collisions between particles will shift the energy of a
state with a given number of atoms and molecules in a manner that depends
nonlinearly on the numbers, making it difficult to maintain exact
resonance. Furthermore, once a the system has Rabi flopped to molecules,
the depopulated atomic condensate makes Bose stimulation, and thus neglect
of the $\k\neq 0$ BEC modes, questionable. And so we investigate the more
robust possibility of rapid adiabatic passage~\cite{ALL87}. From the
Hamiltonian~\eq{SIMP_HAM}, it should be clear that the system will favor
all atoms for large positive detunings, while favoring all molecules for
large negative detunings. With $\Omega$ established as the characteristic
frequency for collective atom-molecule conversion, changes in the
detuning that are of the order of
$\Omega$, and occur over a time of order $\Omega^{-1}$, should
qualify as adiabatic. Hence, if the detuning is swept in a
linear fashion according to $\delta=-\xi\Omega^2 t$, then dimensionless
sweep rates $\xi\lsimeq 1$ should enable rapid adiabatic passage to a
degenerate gas of Fermi molecules.

\begin{figure}
\centering
\epsfig{file=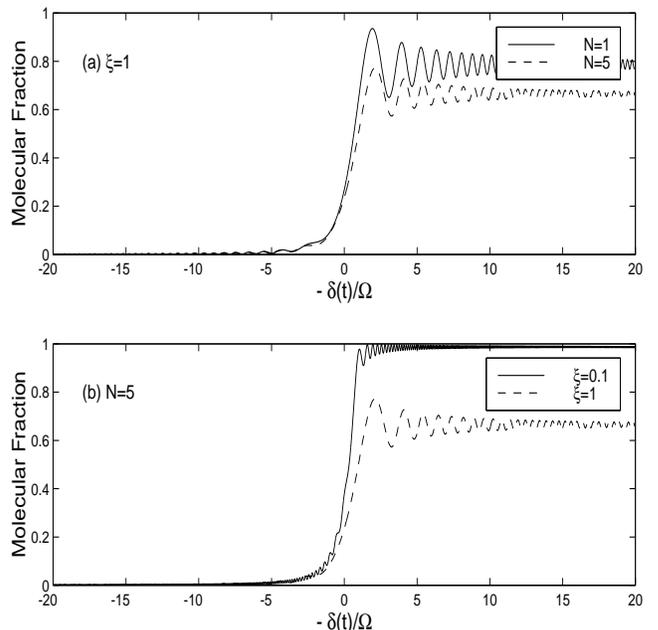,width=8.5cm,height=8.5cm}
\caption{Rapid adiabatic passage from a Fermi-Bose gas of atoms to
Fermi molecules. The detuning is swept
as $\delta(t)=-\xi\Omega^2t$, and $N_B=N_F=N$. (a) For borderline
adiabaticity, 
$\xi=1$, increasing the
number of initial atoms from $N=1$ to $N=5$ indicates that many-body
effects reduce the efficiency. (b) For $N=5$ and $\xi=0.1$,
near-unit conversion is still possible, despite many-body effects.}
\label{SWEEP}
\end{figure}

Our suspicions are again
corroborated by a Fock-state-based numerical solution to the Schr\"odinger
equation, shown in Fig.~\ref{SWEEP}. While many-body effects appear to
rather seriously affect the efficiency of a marginally adiabatic sweep
($\xi=1$) compared to the $N=1$ case, the difference between $N=5$ and
$N=10$ (not shown) is in fact small. Overall, many-body effects are
expected to be weak for near-adiabatic sweeps ($\xi\sim 1$), and
vanishing for sweeps that are deep-adiabatic ($\xi\lsimeq 0.1$), again in
agreement with our BEC experience~\cite{JAV99}. On the matter of
photodissociation to noncondensate modes--and the related pair
correlations\cite{KOK02,MAC02,KOE03}--we note that, in the all-boson
case~\cite{ROGUE}, such transitions can be neglected for a sweep directed
as in Fig.~\ref{SWEEP}, i.e., for
$\dot\delta<0$. Given the success of the analogy so far, similar results
are expected for a Fermi system.

Before closing, we estimate some explicit numbers. We eschew
photoassociation because of the losses associated with the
electronically-excited state~\cite{TWO_CO}, and focus on the atom-molecule
coupling provided by the \rbpot Feshbach resonance located at
$B_0=534\,$G~\cite{SIM03}, which has a width $\Delta_R=4\,$G and an
associated zero-field Fermi-Bose-atom scattering length
$a_{aa}^{FB}=-17.8\,$nm. Accordingly, the atom-molecule coupling is
${\cal K}\approx (4\pi|a_{aa}^{FB}|
  \mu_{\rm Bohr}\Delta_R/m_r)^{1/2}
    =0.14\,\rm{cm}^{3/2}\,\rm{s}^{-1}$,
where we have estimated the difference between the Fermi-Bose atom pair
and molecular magnetic moments to be equal to the Bohr magneton
$\mu_{\rm Bohr}$.
Assuming $N_B=10^5$ condensate atoms in a trap with respective radial and
axial frequencies $\omega_r/2\pi=215\,$Hz and
$\omega_a/2\pi=16.3\,$Hz~\cite{MOD02}, the density of bosons is
$\rho_B=8.1\times 10^{13}\,\rm{cm}^{-3}$. As for the fermions, we assume
a modest number, say, $N_F=10^3$, which has three consequences: (i)  the
atomic BEC will act as a reservoir, thus absorbing any heat created by
holes in the Fermi sea~\cite{TIM01b}; (ii) barring an unfortunately large
scattering length for Bose-atom and Fermi-molecule collisions, we can
neglect the possibility of any Fermi-Bose collapse
instabilities~\cite{MOD02}; (iii) the size of the Fermi cloud
($R_F=8.3\,\mu\rm{m}$) is smaller than the BEC ($R_B=10\,\mu\rm{m}$), so
that overlap is not an issue. Moreover, for
$N_B\gg N_F$, the timescale for atom-molecule conversion is
$\tau_{a2m}\sim (\sqrt{\rho_B}\,\K)^{-1}=8.2\times 10^{-7}\,\rm{s}$.
This timescale is safely below the fastest timescale for trapped-atom
motion
$\tau_t=(\omega_r/2\pi)^{-1}=4.7\times 10^{-3}\,\rm{s}$,
justifying our neglect of trap dynamics and the kinetic
energy; physically put, this means that the Fermi energy is negligible
compared to the atom-molecule coupling strength.

Lastly we discuss the neglected role of particle-particle interactions,
i.e., collisions. These are described in terms of a single
parameter, the
$s$-wave scattering length $a$, where the value of $a$
differs depending on whether one has in mind Bose-Bose atom,
Fermi-Bose atom, Fermi-Bose atom-molecule, or Fermi-Fermi atom-molecule
collisions ($s$-wave Fermi-Fermi atom and molecule collisions are of
course Pauli blocked). Whatever the species involved, the real
part of $a$ describes elastic collisions between particles, and leads to
so-called mean-field shifts.  The imaginary part need only be considered
for atom-molecule collisions terms, and arises because the molecules are
vibrationally very hot, and therefore undergo vibrational relaxation
induced by collisions with unconverted atoms. The collisional coupling
strength is then
$\lambda=2\pi\hbar a/m^*$, where $m^*$ is the mass of the
atom or the reduced mass of the atom-atom (atom-molecule) pair, and it is
included in the theory as a density dependent detuning. The units of the
collisional coupling are
$[\lambda]=\rm{cm}^3/\rm{s}$, so that an appropriate density factor is
needed to arrive at a collisional rate. Estimating the real part of
$a$ from the \rbpot collisions\cite{SIM03}, and the imaginary part from
collisions~\cite{vABE99,YUR99}, a typical density
($\rho\sim10^{14}$) leads to elastic and inelastic collisions that
occur on a timescale
$\tau_c\sim(\rho\lambda)^{-1}\gsimeq1\,\rm{ms}$. For rapid
adiabatic passage, the contribution from the collisional interaction is
negligible compared to the detuning, except near-resonance; however, the
system only spends about $\tau_{a2m}\sim1\rm{\mu s}$ in this region, which
is short enough to expect collisions to be negligible. If, by chance,
collisions do turn out to be a problem, the trap could be expanded to
further reduce the rate of collisions with unconverted atoms.

In conclusion, we highlight
that the term $\Omega=\sqrt{N}\,\kappa$ was previously
referred to as the Bose-enhanced free-bound coupling, and the detuning
sweep, for example, referred to as Bose-stimulated rapid adiabatic passage
from atoms to molecules. However, this behavior is now clearly independent
of statistics, so that Bose stimulation
of free-bound association has nothing whatsoever to do with Bose
statistics, but is instead a many-body cooperative effect that applies
equally well to Fermi-degenerate systems. We expect
this situation to arise whenever the system is addressed as a unit,
i.e., when the atom-molecule coupling strength is larger than the Fermi
energy. To four-wave mixing in a Fermi gas of atoms~\cite{MOO01,KETT01},
we therefore add a further example of fermions mimicking an effect that
was previously attributed to Bose-Einstein statistics. Such behavior
should allow for a shortcut to a degenerate gas of Fermi molecules via a
collective photoassociation or Feshbach resonance.

The authors acknowledge Juha Javanainen for helpful
discussions. This work supported by the Academy
of Finland (MM and KAS, project 50314), and the Magnus Ehrnrooth
foundation (OD).


\vspace{-0.5cm}


\begin{references}

\bibitem{DEM99}
B. DeMarco and D. S. Jin, Science {\bf 285}, 1703 (1999).

\bibitem{WEI98}
  J. D. Weinstein \etal, Nature {\bf 395}, 148 (1998).

\bibitem{BET00}
  H. L. Bethlem \etal, Nature {\bf 406}, 491 (2000).

\bibitem{WEI99}
  J. Weiner, V. S. Bagnato, S. Zilio, and P. S. Julienne,
    \rmp {\bf 71}, 1 (1999).

\bibitem{SCH02}
  U. Schl\"oder, C. Silber, T. Deuschle, and C. Zimmermann,
    Phys. Rev. A {\bf 66}, 061403 (2002).

\bibitem{WYN00}
R. Wynar \etal, Science {\bf 287}, 1016 (2000).

\bibitem{GER01}
 J. M. Gerton, B. J. Frew, and R. G. Hulet,
  \pra {\bf 64}, 053410 (2001).

\bibitem{MCK02}
C. McKenzie \etal, \prl {\bf 88}, 120403 (2002).

\bibitem{PIC02}
  M. Pichler and R. Hulet, (private communication).

\bibitem{TIM99a}
  E. Timmermans, P. Tommasini, M. Hussein, and A. Kerman,
    Phys. Rep. {\bf 315}, 199 (1999).

\bibitem{COR00} S. L. Cornish \etal, \prl {\bf 85},
1795 (2000).

\bibitem{WEB02}
  T. Weber et al., Science {\bf 299}, 292 (2002).

\bibitem{DON02}
E. A. Donley \etal, Nature (London) {\bf 417},  529
(2002).

\bibitem{TRU01}
  A. G. Truscott \etal, Science {\bf 291}, 2570 (2001).

\bibitem{SCH01}
  F. Schreck \etal, \pra {\bf 64}, 011402 (R) (2001).

\bibitem{MOD02}
  G. Modugno \etal, Science {\bf 287}, 2240 (2002).

\bibitem{MAC00}
  For a review of the role of Bose enhancement in photoassociation,
    see M. Mackie and J. Javanainen, \jmo {\bf 47}, 2645 (2000).

\bibitem{JAV02}
  For a brief review of cooperative fermion behavior in four-wave mixing,
    see J.~Javanainen, Nature {\bf 412}, 689 (2002).

\bibitem{JAV99}
  J. Javanainen and M. Mackie, \pra {\bf 59}, R3186 (1999).

\bibitem{HEI00}
  D. J. Heinzen, R. Wynar,  P. D. Drummond, and K. V. Kheruntsyan,  
    \prl {\bf 84}, 5029 (2000).

\bibitem{VAR01}
A. Vardi, V. A. Yurovsky, and J. R. Anglin, \pra {\bf 64}, 063611 (2001).

\bibitem{vABE99}
  F. A. van Abeelen and B. J. Verhaar, \prl {\bf 83}, 1550 (1999).

\bibitem{TIM99}
  E. Timmermans, P. Tommasini, R. C\^ot\'e, M. Hussein,
    and A. Kerman, \prl {\bf 83}, 2691 (1999).

\bibitem{YUR99}
  V. A. Yurovsky, A. Ben-Reuven, P. S. Julienne, and C. J. Williams,
    \pra {\bf 60}, R765 (1999).

\bibitem{ALL87}
  L. Allen and J. H. Eberly, {\it Optical Resonance and Two-Level Atoms},
(Dover, New York, 1987).

\bibitem{KOK02}
  S. J. J. M. F. Kokkelmans and M. J. Holland,
    \prl {\bf 89}, 180401 (2002).

\bibitem{MAC02}
M. Mackie, K.-A. Suominen, and J. Javanainen \prl {\bf 89}, 180403 (2002).

\bibitem{KOE03}
T.~K\"ohler, T. Gasenzer, and K. Burnett, \pra {\bf 67}, 013601 (2003).

\bibitem{ROGUE}
  V. A. Yurovsky and A. Ben-Reuven, \pra \vol{67}, 043611 (2003);
  M. Mackie \etal, (cond-mat/0210131).

\bibitem{TWO_CO}
  A two-color photoassociation scheme, whereby electronically excited
molecules are transferred to a stable molecule state, could of course
avert such losses, and the work herein indicates that ``Bose-stimulated"
Raman adiabatic passage [M. Mackie \etal,
\prl \vol{84}, 3803 (2000)] will in principle create for Fermi molecules
[M. Mackie \etal, (physics/0305057)].

\bibitem{SIM03}
  A. Simoni \etal, \prl \vol{90}, 163202 (2003).

\bibitem{TIM01b}
  E. Timmermans, \prl \vol{87}, 240403 (2001).

\bibitem{MOO01}
  M. G. Moore and P. Meystre, \prl {\bf 86}, 4199 (2001).

\bibitem{KETT01}
  W. Ketterle and S. Inouye, \prl {\bf 86}, 4203 (2001).

\end{references}
\end{document}